\documentclass[twocolumn,twoside,showkeys,preprintnumbers,amsmath,amssymb,secnumarabic,nofootinbib, nobibnotes,epl]{revtex4}
\usepackage{epsfig}
\usepackage{graphicx}
\usepackage{fancyhdr}
\usepackage{pslatex}
\def\fq{{f^q}}

\begin{document}
\title{Nonextensive relativistic nuclear and subnuclear equation
of state}
\author{ A. Lavagno, P. Quarati, A.M. Scarfone}
\affiliation{Dipartimento di Fisica, Politecnico di Torino, I-10129
Torino, Italy \\ Istituto Nazionale di Fisica Nucleare (INFN),
Sezione di Torino e di Cagliari, Italy
}

\begin{abstract}
Following the basic prescriptions of the Tsallis' nonextensive
relativistic thermodynamics, we investigate the relevance of
nonextensive statistical effects on the relativistic nuclear and
subnuclear equation of state. In this framework, we study the first
order phase transition from hadronic to quark-gluon plasma phase by
requiring the Gibbs conditions on the global conservation of the
electric and the baryon charges. The relevance of small deviations
from the standard extensive statistics is investigated in the
context of intermediate and high energy heavy-ion collisions.
\end{abstract}


\maketitle
\thispagestyle{fancy}
\setcounter{page}{1}

\section{Introduction}
The determination of the properties of nuclear matter as functions
of densities, temperature and protons fraction is a fundamental
task in nuclear and subnuclear physics. Several experimental
observations and theoretical calculations clearly indicate that
hadrons dissociate into a plasma of their elementary constituents,
quarks and gluons (QGP), at density several times the nuclear
matter density and/or at temperature above few hundreds MeV, which
is the critical temperature $T_c$ of the transition from the QGP
phase to the hadronic gas phase and viceversa. Such a QGP is
expected to have occurred in the early stages of the Universe and
can be found in dense and hot stars, neutron stars,
nucleus-nucleus high energy collisions where heavy ions are
accelerated to relativistic energies~\cite{hwa}. After collision,
a fireball is created which may realize the conditions of the QGP.
The plasma then expands, cools, freezes-out into hadrons, photons,
leptons that are detected and analyzed \cite{biro08}.

It is a rather common opinion that, because of the extreme
conditions of density and temperature in ultrarelativistic heavy
ion collisions, memory effects and long--range color interactions
give rise to the presence of non--Markovian processes in the
kinetic equation affecting the thermalization process toward
equilibrium as well as the standard equilibrium distribution
\cite{hei,biro,albe}. A rigorous determination of the conditions
that produce a nonextensive behavior, due to memory effects and/or
long--range interactions,  should be based on microscopic
calculations relative to the parton plasma originated during the
high energy collisions. At this stage we limit ourselves to
consider the problem from a qualitative point of view on the basis
of the existing theoretical calculations and experimental
evidences.

On the other hand, over the last years, there has been an
increasing evidence that the generalized non-extensive statistical
mechanics, proposed by Tsallis \cite{tsallis,tsamendes,gellmann}
and characterized by a power-law stationary particle distribution,
can be considered as a
 basis for a theoretical framework appropriate to incorporate, at
least to some extent and without going into microscopic dynamical
description, long-range interactions, long-range microscopic
memories and/or fractal space-time constraints. A considerable
variety of physical issues show a quantitative agreement between
experimental data and theoretical analysis based on Tsallis'
thermostatistics. In particular, there is a growing interest in
high energy physics applications of non-extensive statistics
\cite{wilk1,plb2001,rafelski,bediaga,beck,biroprl05}. Several
authors outline the possibility that experimental observations in
relativistic heavy-ion collisions can reflect non-extensive
statistical mechanics effects during the early stage of the
collisions and the thermalization evolution of the system
\cite{albe,biro04,biropla08,wilk2,wilk08,lavaphysa,physicaA2008}.
In this context, it is relevant to observe that the statistical
origin of the nonextensive statistics lies in the deformation of
the Boltzmann entropy. Let us note that in literature other
statistical generalizations are present, such as the $q$-deformed
thermostatistics inspired to $q$-deformed quantum algebra and
quantum groups with a different origin
\cite{pla97,pre2000,epj2006,jpa2008}.

From the above considerations, it appears reasonable that in
regime of high density and temperature both hadron and quark-gluon
Equation Of State (EOS) can be sensibly affected by nonextensive
statistical effects \cite{physicaEOS}. Furthermore, in this
context it is very remarkable to observe that the relevance of
these effects on the relativistic hadronic equation of state has
also been recently investigated in Ref.~\cite{silva}.

The aim of this paper is to study the behavior of the nuclear
equation of state at finite temperature and baryon density and to
explore the existence of a hadron-quark mixed phase at a fixed
value of the proton fraction $Z/A$.

\section{Relativistic nonextensive thermodynamics}

In order to study, from a phenomenological point of view,
experimental observable in relativistic heavy-ion collisions, in
this Section we present the basic macroscopic thermodynamic
variables and kinetic theory in the language of the nonextensive
relativistic kinetic theory.

Let us start by introducing the particle four-flow in the phase
space as \cite{lavapla}
\begin{equation}
N^\mu(x)=\frac{1}{Z_q}\int \frac{d^3p}{p^0} \, p^\mu \,f(x,p) \; ,
\label{nmu}
\end{equation}
and the energy-momentum flow as
\begin{equation}
T^{\mu\nu}(x)=\frac{1}{Z_q}\int \frac{d^3p}{p^0} \, p^\mu p^\nu \,
f^q(x,p) \; , \label{tmunu}
\end{equation}
where we have set $\hbar=c=1$, $x\equiv x^\mu=(t,{\bf x})$,
$p\equiv p^\mu=(p^0,{\bf p})$, $p^0=\sqrt{{\bf p}^2+m^2}$ being
the relativistic energy and $f(x,p)$ the particle distribution
function. In the above ${Z_q}=\int d\Omega \, [f(x,p)]^q $ is the
nonextensive partition function, $d\Omega$ stands for the
corresponding phase space volume element and $q$ is the
deformation parameter. The four-vector $N^\mu=(n,{\bf j})$
represents the probability density $n=n(x)$ (which is normalized
to unity) and the probability flow ${\bf j}={\bf j}(x)$. The
energy-momentum tensor contains the normalized $q$-mean
expectation value of the energy density, as well as the energy
flow, the momentum and the momentum flow per particle. Its
expression follows directly from the definition of the mean
$q$-expectation value in nonextensive statistics \cite{tsamendes};
for this reason $T^{\mu\nu}$ it is given in terms of $f^q(x,p)$.

On the basis of the above definitions, one can show that it is
possible to obtain a generalized non-linear relativistic Boltzmann
equation \cite{lavapla}
\begin{equation}
p^\mu \partial_{\mu}\left[f(x,p)\right]^q=C_q(x,p)  \; ,
\label{boltz}
\end{equation}
where the function $C_q(x,p)$  implicitly defines a generalized
nonextensive collision term
\begin{eqnarray}
C_q(x,p)=&&\frac{1}{2}\! \int\!\!\frac{d^3p_1}{p^0_1}
\frac{d^3p{'}}{p{'}^0} \frac{d^3p{'}_1}{p{'}^0_1} \Big
\{h_q[f{'},f{'}_1]  W(p{'},p{'}_1\vert p,p_1) \nonumber \\
&&-h_q[f,f_1]  W(p,p_1\vert p{'},p{'}_1) \Big\}\; .
\end{eqnarray}
Here $W(p,p_1\vert p{'},p{'}_1)$ is the transition rate between a
two-particle state with initial four-momenta $p$ and $p_1$ and a
final state with four-momenta $p{'}$ and $p{'}_1$; $h_q[f,f_1]$ is
the $q$-correlation function relative to two particles in the same
space-time position but with different four-momenta $p$ and $p_1$,
respectively. Such a transport equation conserves the probability
normalization (number of particles) and is consistent with the
energy-momentum conservation laws in the framework of the
normalized $q$-mean expectation value. Moreover, the collision
term contains a generalized expression of the molecular chaos and
for $q>0$ implies the validity of a generalized $H$-theorem, if
the following, nonextensive, local four-density entropy is assumed
(henceforward we shall set Boltzmann constant $k_{_B}$ to unity)
%
\begin{equation}
S_q^\mu(x)=- \int \frac{d^3p}{p^0} \,p^\mu f[(x,p)]^q [\ln_q
f(x,p)-1] \; , \label{entro4}
\end{equation}
where we have used the definition $\ln_q x=(x^{1-q}-1)/(1-q)$, the
inverse function of the Tsallis' $q$-exponential function
\begin{equation}
e_q(x)=[1+(1-q)x]^{1/(1-q)} \; , \label{tsaexp}
\end{equation}
which satisfies the property $e_q(\ln_q x)=x$.

The above expression is written in a covariant form, in fact
$S^{\mu}_q=(S^0_q,S^i_q)$, with $i=1,2,3$, and correctly
transforms as a four-vector under Lorentz
transformations~\cite{lavapla}, where $S^0_q$ is the
re\-lativistic Tsallis nonextensive local entropy density and
$S^i_q$ is the Tsallis entropy flow. Note that for $q\rightarrow
1$, Eq.(\ref{entro4}) reduces to the well known four-flow entropy
expression \cite{groot}.

At equilibrium, the solution of the above Boltzmann equation is a
relativistic Tsallis-like (power law) distribution and can be
written as
\begin{equation}
f_{eq}(p)= \frac{1}{Z_q}\left [1-(1-q) \frac{p^\mu U_\mu}{T}
\right]^{1/(1-q)} \; , \label{nrdistri}
\end{equation}
where $U_\mu$ is the hydrodynamic four-velocity \cite{groot} and
$f_{eq}$ depends only on the momentum in the absence of an external
field. At this stage, $T$ is a free parameter and only in the
derivation of the equation of state it will be identified with the
physical temperature.

We are able now to evaluate explicitly all other thermodynamic
variables and provide a complete macroscopic description of a
relativistic system at the equilibrium. Considering the
decomposition of the energy-momentum tensor: $T^{\mu\nu}=\epsilon\,
U^\mu U^\nu-P\, \Delta^{\mu\nu}$, where $\epsilon$ is the energy
density, $P$ the pressure and $\Delta^{\mu\nu}=g^{\mu\nu}-U^\mu
U^\nu$, the equilibrium pressure can be calculated as
\begin{equation}
P=-\frac{1}{3} T^{\mu\nu}\Delta_{\mu\nu}=-\frac{1}{3\,Z_q}
\int\frac{d^3p}{p^0} p^\mu p^\nu
\Delta_{\mu\nu}\fq_{\!\!\!\!\!eq}(p) \; .  \label{pressrel}
\end{equation}
Setting $\tau=p^0/T$ and $z=m/T$, the above integral can be
expressed as
\begin{equation}
P=\frac{4\pi}{Z_q}\,m^2\,T^2\, K_2(q,z) \; ,\label{press}
\end{equation}
where we have introduced the $q$-modified Bessel function of the
second kind as follows
\begin{equation}
K_n(q,z)=\frac{2^n n!}{(2n)!}\frac{1}{z^n}\int^\infty_z \!\!d\tau
(\tau^2-z^2)^{n-1/2}\, \left(e^{-\tau}_q\right)^q \;
,\label{qbessel}
\end{equation}
and $e_q(x)$ is the $q$-modified exponential defined in
Eq.(\ref{tsaexp}).

Similarly, the energy density $\epsilon$ can be obtained from the
following expression
\begin{equation}
\epsilon=T^{\mu\nu}U_\mu U_\nu=\frac{1}{Z_q}  \int\frac{d^3p}{p^0}
(p^\mu U_\mu)^2 \fq_{\!\!\!\!\!eq}(p) \; , \label{energyrel}
\end{equation}
and, after performing the integration, it can be cast into the
compact expression:
\begin{equation}
\epsilon=\frac{4\pi}{Z_q}\, m^4 \left [
3\frac{K_2(q,z)}{z^2}+\frac{K_1(q,z)}{z}\right ] \; .
\end{equation}
Thus the energy per particle $e=\epsilon/n$ is
\begin{equation}
e=3 \,T +m \,\frac{K_1(q,z)}{K_2(q,z)} \; ,
\end{equation}
which has the same structure of the relativistic expression
obtained in the framework of the equilibrium Boltzmann-Gibbs
statistics \cite{groot}.

In the non-relativistic limit ($p\ll 1$) the energy per particle
reduces to the well-known expression
\begin{equation}
e\simeq m+\frac{3}{2}\,T  \; ,
\end{equation}
and no explicit $q$-dependence is left over.

Hence from the above results it appears that, in searching for the
relevance of nonextensive statistical effects, both microscopic
observable, such as particle distribution, correlation functions,
fluctuations of thermodynamical variables, and macroscopic
variables, such as energy density or pressure, can be affected by
the deformation parameter $q$.

In this context, it appears relevant to observe that, in
Ref.~\cite{biroprl05} nonextensive Boltzmann equation has been
studied and proposed for describing the hadronization of quark
matter. Moreover, starting from the above generalized relativistic
kinetic equations, in Ref.\cite{wilk08} the authors have recently
formulated a nonextensive hydrodynamic model for multiparticle
production processes in relativistic heavy-ion collisions. These
works represent an important bridge for a close connection between
a microscopic nonextensive model and experimental observable.

Finally, let us remind the reader that for a system of particles in
a degenerate regime the above classical distribution function
(\ref{nrdistri}) has to be modified by including the fermion and
boson quantum statistical prescriptions. For a dilute gas of
particles and/or for small deviations from the standard extensive
statistics ($q\approx 1$) the equilibrium distribution function, in
the grand canonical ensemble, can be written as \cite{buyu}
\begin{equation}
n(k,\mu)=\frac{1} { [1+(q-1)(E(k)-\mu)/T ]^{1/(q-1)} \pm 1}  \, ,
\label{distrifd}
\end{equation}
where the sign $+$ stands for fermions and $-$ for bosons: hence all
previous results can be easily extended to the case of quantum
statistical mechanics.

\section{Nonextensive nuclear equation of state}

The EOS at densities below the saturation density of nuclear
matter $\rho_0 \approx 0.16$ fm$^{-3}$ is relatively well known
due to the large amount of experimental nuclear data available. At
larger densities there are many uncertainties due the lack of
experimental data; the strong repulsion at short distances of
nuclear force makes, in fact, the compression of nuclear matter to
larger densities quite difficult. Otherwise, in relativistic heavy
ion collision the baryon densities can reach values of a few times
$\rho_0$ and the temperature can exceed the deconfinement critical
temperature $T_c \approx 170$ MeV.

As partially discussed in the Introduction, hadronic matter is
expected to undergo a phase transition into a deconfined phase of
quarks and gluons at large densities and/or high temperatures.
However, the extraction of experimental information about the EOS
of matter at large densities and temperatures at intermediate and
high energy heavy-ion collisions is very complicated. Possible
indirect indications of a softening of the EOS at the energies
reached at AGS have been discussed several times in the literature
\cite{stoecker,prl}. In particular, a recent analysis
\cite{ivanov} based on a 3-fluid dynamics simulation suggests a
progressive softening of the EOS tested through heavy-ion
collisions at energies ranging from 2A GeV up to 8A GeV. On the
other hand, the information coming from experiments with
heavy-ions at intermediate and high energy collisions is that, for
symmetric or nearly symmetric nuclear matter, the critical density
(at low temperatures) appears to be considerably larger than
nuclear matter saturation density $\rho_0$. Concerning
non-symmetric matter, general arguments based on Pauli principle
suggest that the critical density decreases with $Z/A$. Therefore,
the transition's critical densities are expected to sensibly
depend on the isospin of the system \cite{ditoro}. Moreover, the
analysis of observations of neutron stars, which are composed of
$\beta$-stable matter for which $Z/A\le 0.1$ (the matter
constituting neutron stars is strongly isospin asymmetric, being
composed of a large amount of neutrons and a small fraction of
protons) can also provide hints on the structure of extremely
asymmetric matter at high density. No data on the quark
deconfinement transition are at the moment available for
intermediate values of $Z/A$. Recently, it has been proposed by
several groups to produce unstable neutron-rich beams at
intermediate energies. These new experiments can open the
possibility to explore in laboratory the isospin dependence of the
critical densities.

The scenario we are going to explore in this last Section
corresponds to the situation realized in experiments at not too
high energy. In this condition, only a small fraction of
strangeness can be produced and, therefore, we limit ourselves to
study the deconfinement transition from nucleonic matter into up
and down quark matter. In the next two subsections, we will study
the two corresponding EOSs separately, on the basis of the
previously reported nonextensive relativistic thermodynamic
relations. The existence of the hadron-quark mixed phase will be
studied in the third subsection. This investigation may be helpful
also in view of the future experiments planned, e.g., at the
facility FAIR at GSI \cite{senger}.

\subsection{Nonextensive hadronic equation of state}

The relativistic, field theoretical approach to nuclear EOS was
used first by Walecka and Boguta-Bodmer in the mid-1970s
\cite{walecka,boguta}. This theory describes the interaction
between nucleons through the exchange of two mesons, the scalar
field $\sigma$ and the vector field $\omega$.  The model of
Walecka has two free parameters: the two ratios between the
nucleon-meson coupling constants and the masses of the mesons. The
saturation density and binding energy per nucleon (calculated at
the the saturation density) of nuclear matter can be fitted
exactly in the simplest version of this model but other properties
of nuclear matter, as e.g. incompressibility, cannot be
reproduced. To overcome these difficulties, the model has been
modified introducing in the Lagrangian two terms of
self-interaction for the $\sigma$ which are crucial to reproduce
the empirical incompressibility of nuclear matter and the
effective mass of nucleons $M^*$ (again calculated at the
saturation density). Moreover, the introduction of an isovector
meson $\rho$ allows to reproduce the correct value of the
empirical symmetry energy \cite{serot,glen}.

In the following, we will use a relativistic mean field
self-consistent theory of nuclear matter in which nucleons
interact through the nuclear force mediated by the exchange of
virtual isoscalar and isovector mesons ($\sigma,\omega,\rho$) with
a Lagrangian density \cite{glen}
\begin{eqnarray}\label{eq:1}
{\cal L } &=& \bar{\psi}[i\gamma_{\mu}\partial^{\mu}-(M-
g_{\sigma}\sigma)
-g{_\omega}\gamma_\mu\omega^{\mu}-g_\rho\gamma^{\mu}\vec\tau\cdot
\vec{\rho}_{\mu}]\psi \nonumber \\&&
+\frac{1}{2}(\partial_{\mu}\sigma\partial^{\mu}\sigma-m_{\sigma}^2\sigma^2)
-U(\sigma)+\frac{1}{2}m^2_{\omega}\omega_{\mu} \omega^{\mu}
\nonumber
\\&&
+\frac{1}{2}m^2_{\rho}\vec{\rho}_{\mu}\cdot\vec{\rho}^{\mu}
-\frac{1}{4}F_{\mu\nu}F^{\mu\nu}
-\frac{1}{4}\vec{G}_{\mu\nu}\vec{G}^{\mu\nu}\,,
\end{eqnarray}
where $M=939$ MeV is the vacuum baryon mass,
($\sigma$,$\omega_{\mu}$) are the isoscalar (scalar,vector) meson
fields, while $\vec{\rho}_{\mu}$ is the corresponding isovector
ones. The field strength tensors for the vector mesons are given
by the usual expressions
$F_{\mu\nu}\equiv\partial_{\mu}\omega_{\nu}-\partial_{\nu}\omega_{\mu}$,
$\vec{G}_{\mu\nu}\equiv\partial_{\mu}\vec{\rho}_{\nu}-\partial_{\nu}\vec{\rho}_{\mu}$,
and the $U(\sigma)$ is a nonlinear potential of $\sigma$ meson
\begin{eqnarray}
U(\sigma)=\frac{1}{3}a\sigma^{3}+\frac{1}{4}b\sigma^{4}\,.
\end{eqnarray}
This last term is usually introduced to achieve a reasonable
compression modulus for equilibrium nuclear matter.

The field equations in a mean field approximation are
\begin{eqnarray}\label{eq:2}
&&(i\gamma_{\mu}\partial^{\mu}-(M- g_{\sigma}\sigma)-
g_\omega\gamma^{0}{\omega_0}-g_\rho\gamma^{0}{\tau_3}{\rho_0})\psi=0\,,\nonumber \\
&&m_{\sigma}^2\sigma+ a{{\sigma}^2}+ b{{\sigma}^3}=g_\sigma<\bar\psi\psi>=g_\sigma{\rho}_S\,, \nonumber \\
&&m^2_{\omega}\omega_{0}=g_\omega<\bar\psi{\gamma^0}\psi>=g_\omega\rho_B\,,
\nonumber \\
&&m^2_{\rho}\rho_{0}=g_\rho<\bar\psi{\gamma^0}\tau_3\psi>=g_\rho\rho_I\,,
\end{eqnarray}
where $\rho_I=\rho_p-\rho_n$ is the isospin density, $\rho_B$ and
$\rho_S$ are the baryon and the scalar densities, respectively.
They are given by
\begin{eqnarray}
&&\rho_{B}=2 \sum_{i=n,p} \int\frac{{\rm
d}^3k}{(2\pi)^3}[n_i(k)-\overline{n}_i(k)]\,, \nonumber\\
&&\rho_S=2 \sum_{i=n,p} \int\frac{{\rm
d}^3k}{(2\pi)^3}\,\frac{M_i^\star}{E_i^\star}\,
[n_i^q(k)+\overline{n}_i^q(k)]\,,
\end{eqnarray}
where $n_i(k)$ and $\overline{n}_i(k)$ are the $q$-deformed
fermion particle and antiparticle distributions given in
Eq.(\ref{distrifd}); more explicitly
\begin{eqnarray}
n_i(k)=\frac{1} { [1+(q-1)(E_i^\star(k)-\mu_i^\star)/T
]^{1/(q-1)} + 1} \, , \\
\overline{n}_i(k)=\frac{1} {[1+(q-1)(E_i^\star(k)+\mu_i^\star)/T
]^{1/(q-1)} + 1} \, .
\end{eqnarray}

The nucleon effective energy is defined as
${E_i}^\star(k)=\sqrt{k^2+{{M_i}^\star}^2}$, where
${M_i}^\star=M_{i}-g_\sigma \sigma$. The effective chemical
potentials $\mu_i^\star$  are given in terms of the vector meson
mean fields $\mu_i^\star={\mu_i}-g_\omega\omega_0 \mp
g_{\rho}\rho_0$ ($-$ proton, $+$ neutron), where $\mu_i$ are the
thermodynamical chemical potentials
$\mu_i=\partial\epsilon/\partial\rho_i$. At zero temperature they
reduce to the Fermi energies $E_{Fi} \equiv
\sqrt{k_{Fi}^2+{M_i^\star}^2}$ and the nonextensive statistical
effects disappear. The meson fields ($\sigma$, $\omega_0$ and
$\rho_0$) are obtained as a solution of the field equations in
mean field approximation and the related couplings ($g_\sigma$,
$g_\omega$ and $g_\rho$) are the free parameter of the model
\cite{glen}.

On the basis of Eqs.(\ref{tmunu}), (\ref{pressrel}) and
(\ref{energyrel}), the pressure and the energy density can be
written as
\begin{eqnarray}
P&=&\frac{2}{3} \sum_{i=n,p} \int \frac{{\rm d}^3k}{(2\pi)^3}
\frac{k^2}{E_{i}^\star(k)} [n_i^q(k)+\overline{n}_i^q(k)]
 \nonumber \\
&-&\frac{1}{2}m_\sigma^2\sigma^2 - U(\sigma)+
\frac{1}{2}m_\omega^2\omega_0^2+\frac{1}{2}m_{\rho}^2 \rho_0^2\,,\\
\epsilon&=& {2}\sum_{i=n,p}\int \frac{{\rm
d}^3k}{(2\pi)^3}E_{i}^\star(k)
[n_i^q(k)+\overline{n}_i^q(k)]\nonumber \\
&+&\frac{1}{2}m_\sigma^2\sigma^2+U(\sigma)
+\frac{1}{2}m_\omega^2\omega_0^2+\frac{1}{2}m_{\rho}^2 \rho_0^2\,,
\end{eqnarray}

Note that statistical mechanics enters as an external ingredient
in the functional form of the "free" particle distribution of
Eq.~(\ref{distrifd}). Since all the equations must be solved in a
self-consistent way, the presence of nonextensive statistical
effects in the particle distribution function influences the
many-body interaction in the mean field self-consistent solutions
obtained for the meson fields.

\begin{figure}
\resizebox{0.48\textwidth}{!}{%
\includegraphics*[90,550][500,800]{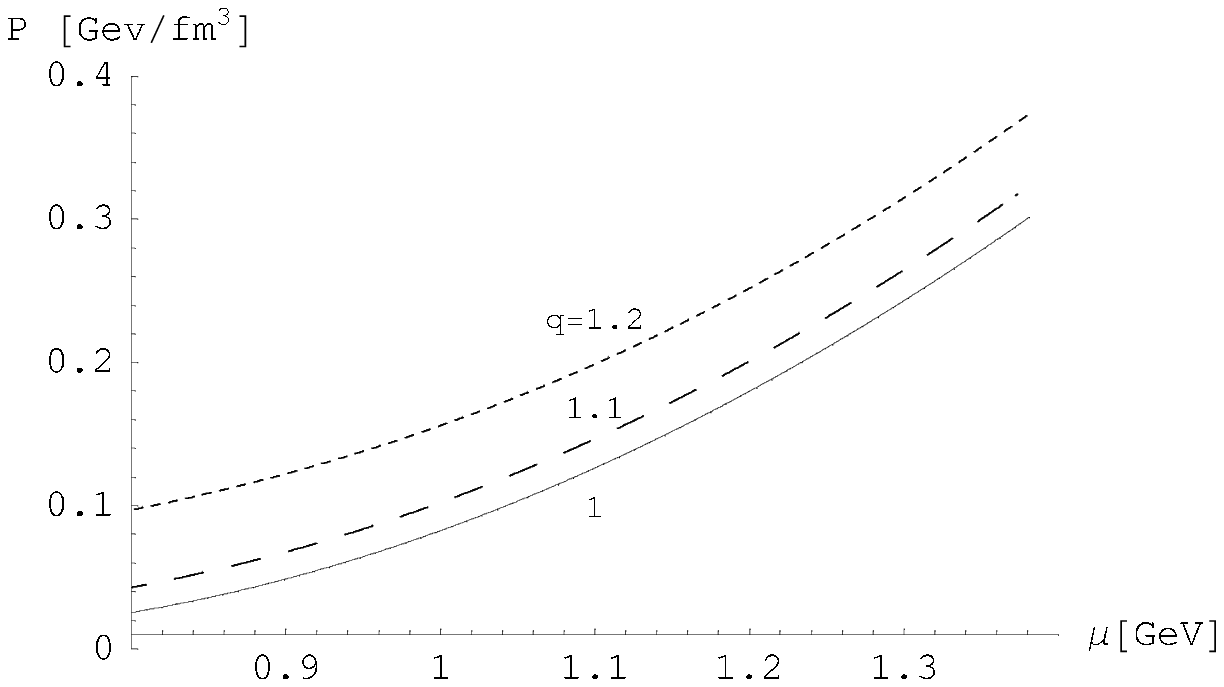}
} \caption{Hadronic equation of state: pressure versus baryon
chemical potential for different values of $q$. In the figure
$T=100$ MeV and $Z/A=0.4$.} \label{figpmuh}
\end{figure}

In Fig. \ref{figpmuh}, we report the resulting hadronic EOS:
pressure as a function of baryon chemical potential for different
values of $q$. Since in the previous investigations we have
phenomenologically obtained values of $q$ greater than unity
\cite{albe,physicaA2008} , we will concentrate our analysis to
$q>1$. The results are plotted at the temperature $T=100$ MeV, at
fixed value of $Z/A=0.4$ and we have used the GM2 set of
parameters of Ref.\cite{glen}. The range of the considered baryon
density and the chosen values of the parameters correspond to a
physical situation which can be realized in the recently proposed
high energy heavy-ion collisions experiment at GSI~\cite{gsi}.

\subsection{Nonextensive QGP equation of state}

In high density hadronic matter, baryons are forced to stay so
close one to another that they would overlap. At such densities,
constituent quarks are shared by neighboring baryons and should
eventually become mobile over a distance larger than the typical
size of one baryon. This means that quarks become deconfined and
that at large densities and/or high temperatures they are the real
degrees of freedom of strongly interacting matter instead of
baryons. The process of deconfinement and the EOS of quark matter
can in principle be described by quantum chromodynamics. However,
in energy density range reached in the core of compact stars and
in relativistic heavy-ion collisions, non-perturbative effects in
the complex theory of QCD are present. For these reasons, simple
phenomenological models are usually adopted to describe quark
matter \cite{mit,schertler,dlplb}.

In this study, due to its simplicity, we use the MIT bag model
\cite{mit}. In this model, quark matter is described as a gas of
free quarks with massless up and down quarks. All the
non-perturbative effects are simulated by the bag constant $B$
which represents the pressure of the vacuum. Following this line,
the pressure, energy density and baryon number density for a
relativistic Fermi gas of quarks in the framework of nonextensive
statistics (see Eqs.(\ref{nmu}), (\ref{tmunu}), (\ref{pressrel})
and (\ref{energyrel})) can be written, respectively, as
\begin{eqnarray}
&P& =\frac{\gamma_f}{3} \sum_{f=u,d} \int^\infty_0 \frac{{\rm
d}^3k}{(2\pi)^3} \,\frac{k^2}{e_f}\,
[n_f^q(k)+\overline{n}_f^q(k)]
-B\,, \label{bag-pressure}\\
&\epsilon& =\gamma_f \sum_{f=u,d}  \int^\infty_0 \frac{{\rm
d}^3k}{(2\pi)^3} \,e_f\, [n_f^q(k)+\overline{n}_f^q(k)]
\label{bag-energy}
+B\,, \\
&\rho& =\frac{\gamma_f}{3} \sum_{f=u,d} \int^\infty_0 \frac{{\rm
d}^3k}{(2\pi)^3}  \,[n_f(k)-\overline{n}_f(k)]\, ,
\label{bag-density}
\end{eqnarray}
where the quark degeneracy for each flavor is $\gamma_f=6$,
$e_f=(k^2+m_f^2)^{1/2}$, $n_f(k)$ and $\overline{n}_f(k)$ are the
$q$-deformed particle and antiparticle quark distributions
\begin{eqnarray}
n_f(k)=\frac{1} { [1+(q-1)(e_f(k)-\mu_f)/T
]^{1/(q-1)} + 1} \, , \\
\overline{n}_f(k)=\frac{1}{[1+(q-1)(e_f(k)+\mu_f)/T ]^{1/(q-1)} +
1} \, .
\end{eqnarray}

Similar expressions for the pressure and the energy density can be
written for the gluons treating them as a massless $q$-deformed
Bose gas with zero chemical potential. Explicitly, we can
calculate the nonextensive pressure $P_g$ and density energy
$\epsilon_g$ for gluons as
\begin{eqnarray}
&P_g& =\frac{\gamma_g}{3} \int^\infty_0 \frac{{\rm
d}^3k}{(2\pi)^3}
\,\frac{k}{[1+(q-1)\,k/T]^{q/(q-1)} - 1}\,, \label{gluon-press}\\
&\epsilon_g& =3\, P_g \, , \label{gluon-energy}
\end{eqnarray}
with the gluon degeneracy factor $\gamma_g=16$. Let us observe
that, in the limit $q\rightarrow 1$, we recover the usual
analytical expression for the pressure of gluons:
$P_g=8\pi^2/45\,T^4$.

Since one has to employ the fermion (boson) nonexten\-sive
distributions, the results are not analytical, even in the
massless quark approximation. Hence a numerical evaluations of the
integrals in Eq.s~(\ref{bag-pressure})--(\ref{bag-density}) and
Eq.s~(\ref{gluon-press})--(\ref{gluon-energy}) must be performed.
Let us remember that a similar calculation, only for the
quark-gluon phase, was also performed in Ref.\cite{miller} by
studying the phase transition diagram.

As previously discussed, in this investigation we are limiting our
study to the two-flavor ($f=u, \, d$) massless quarks. As already
remarked, this appears rather well justified for the application
to heavy ion collisions at relativistic (but not
ultra-relativistic) energies, the fraction of strangeness produced
at these energies being small \cite{ditoro2,fuchs}.

\begin{figure}
\resizebox{0.48\textwidth}{!}{%
  \includegraphics*[90,550][500,800]{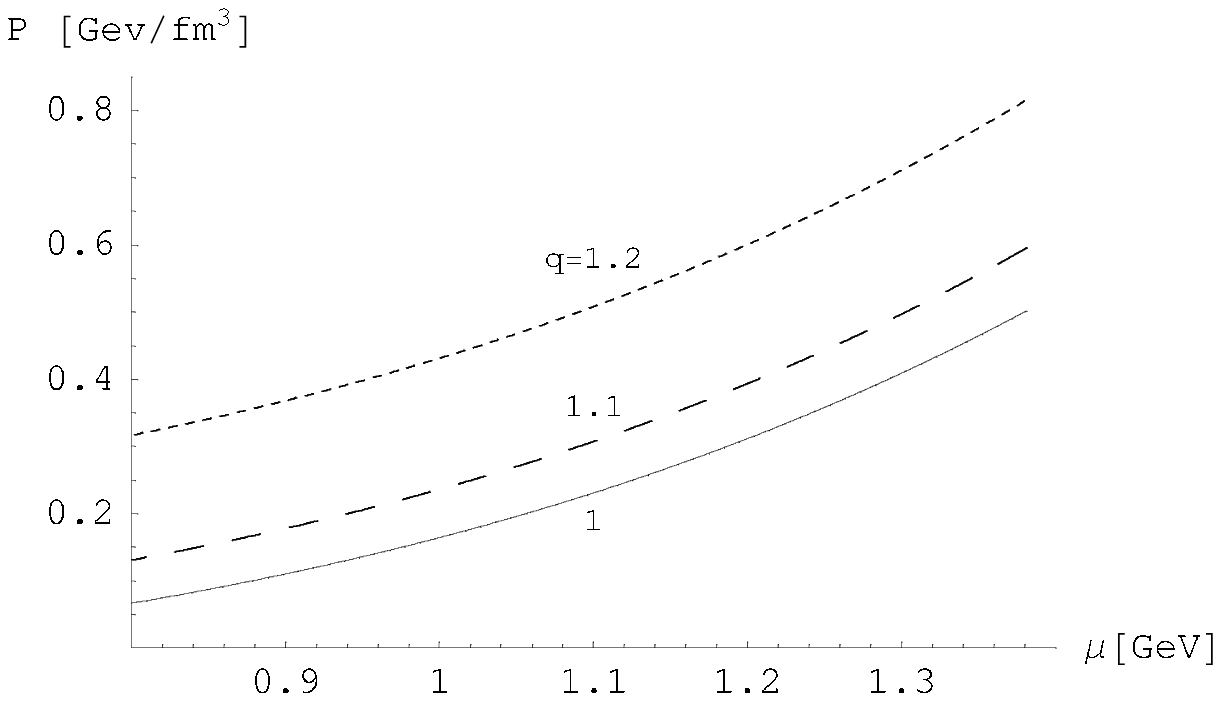}
} \caption{The same as in Fig.~\ref{figpmuh} for the case of the
quark-gluon equation of state.} \label{figpmuq}
\end{figure}

In Fig. \ref{figpmuq}, we report the pressure as a function of the
baryon chemical potential for massless quarks $u$, $d$ and gluons,
for different values of $q$. As in Fig.~\ref{figpmuh}, the results
are plotted at the temperature $T=100$ MeV and at a fixed value of
$Z/A=0.4$; the bag parameter is $B^{1/4}$=170 MeV. In both figures
\ref{figpmuh} and \ref{figpmuq} one can observe sizable effects in
the behavior of the EOS even for small deviations from the
standard statistics.

\subsection{Mixed hadron-quark phase}

In this subsection we investigate the hadron-quark phase transition
at finite temperature and baryon chemical potential by means of the
previous relativistic EOSs. Lattice calculations predict a critical
phase transition temperature $T_c$ of about 170 MeV, corresponding
to a critical energy density $\epsilon_c\approx$ 1 GeV/fm$^3$
\cite{hwa}. In a theory with only gluons and no quarks, the
transition turns out to be of first order. In nature, since the $u$
and $d$ quarks have a small mass, while the strange quark has a
somewhat larger mass, the phase transition is predicted to be a
smooth cross over. However, since it occurs over a very narrow range
of temperatures, the transition, for several practical purposes, can
still be considered of first order. Indeed the lattice data with 2
or 3 dynamical flavours are not precise enough to unambigously
control the difference between the two situations. Thus, by
considering the deconfinement transition at finite density as a the
first order one, a mixed phase can be formed, which is typically
described using the two separate equations of state, one for the
hadronic and one for the quark phase.

To describe the mixed phase we use the Gibbs formalism, which in
Ref.~\cite{glenprd}  has been applied to systems where more than one
conserved charge is present. In this contribution we are studying
the formation of a mixed phase in which both baryon number and
isospin charge are preserved. The main result of this formalism is
that, at variance with the so-called Maxwell construction, the
pressure in the mixed phase is not constant and therefore the
nuclear incompressibility does not vanish. It is important to notice
that from the viewpoint of Ehrenfest's definition, a phase
transition with two conserved charges is considered, in the
literature, not of first, but of second order \cite{muller}.

The structure of the mixed phase is obtained by imposing the Gibbs
conditions for chemical potentials and pressure and by requiring
the global conservation of the total baryon (B) and isospin
densities (I) in the hadronic phase (H) and in the quark phase (Q)
\begin{eqnarray}
&&\mu_B^{(H)} = \mu_B^{(Q)}\, ,\nonumber \\
&&\mu_I^{(H)} = \mu_I^{(Q)}\, ,
\nonumber \\
&& P^{(H)} (T,\mu_{B,I}^{(H)}) =  P^{(Q)} (T,\mu_{B,I}^{(Q)})\, ,
\nonumber \\
&&\rho_B=(1-\chi)\rho_B^H+\chi \rho_B^Q\,, \nonumber \\
&&\rho_I=(1-\chi)\rho_I^H+\chi \rho_I^Q \, ,
\end{eqnarray}
where $\chi$ is the fraction of quark matter in the mixed phase.
In this way we can obtain the binodal surface which gives the
phase coexistence region in the $(T,\rho_B,\rho_I)$ space. For a
fixed value of the conserved charge $\rho_I$, related to the
proton fraction $Z/A \equiv (1+\rho_I/\rho_B)/2$, we study the
boundaries of the mixed phase region in the $(T,\rho_B)$ plane. We
are particularly interested in the lower baryon density (or baryon
chemical potential) border, i.e. the critical/transition density
$\rho_{cr}$, in order to check the possibility of reaching such
$(T,\rho_{cr},\rho_I)$ conditions in a transient state during a
heavy-ion collision at relativistic energies.

\begin{figure}
\resizebox{0.48\textwidth}{!}{%
  \includegraphics*[90,550][500,800]{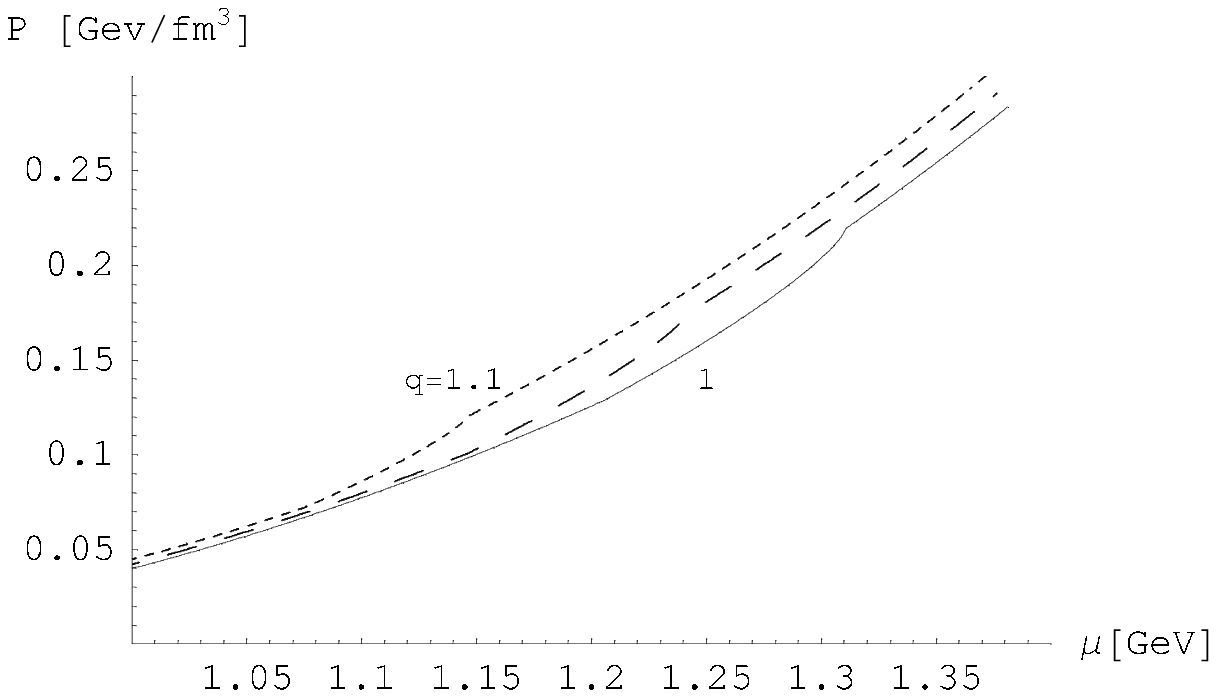}
} \caption{Pressure versus baryon chemical potential in the mixed
hadron-quark phase for different values of $q$. The long-dashed
line corresponds to $q=1.05$. The temperature is fixed a T=60 MeV
and the proton fraction $Z/A=0.4$. } \label{fig_MP1}
\end{figure}

\begin{figure}
\resizebox{0.48\textwidth}{!}{%
  \includegraphics*[90,550][500,800]{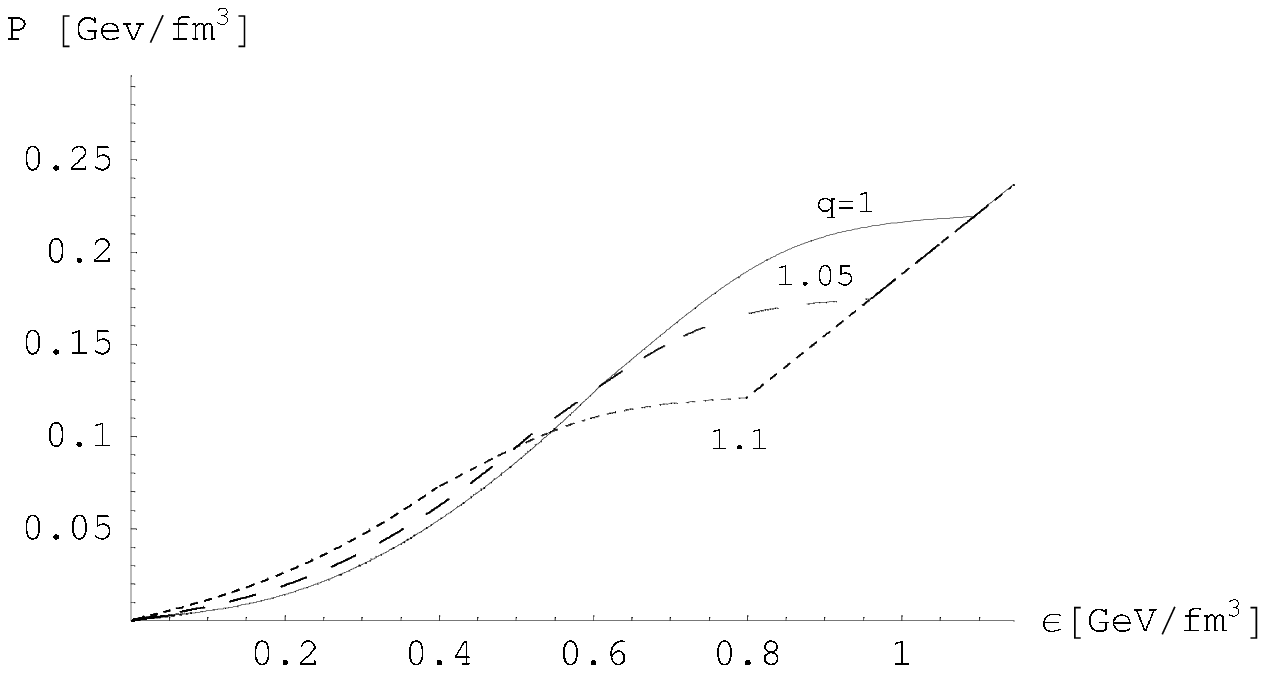}
} \caption{Pressure versus energy density in the mixed
hadron-quark phase for different values of $q$. The temperature is
fixed a T=60 MeV and the proton fraction $Z/A=0.4$.}
\label{fig_MP2}
\end{figure}

\begin{figure}
\resizebox{0.48\textwidth}{!}{%
  \includegraphics*[90,550][500,800]{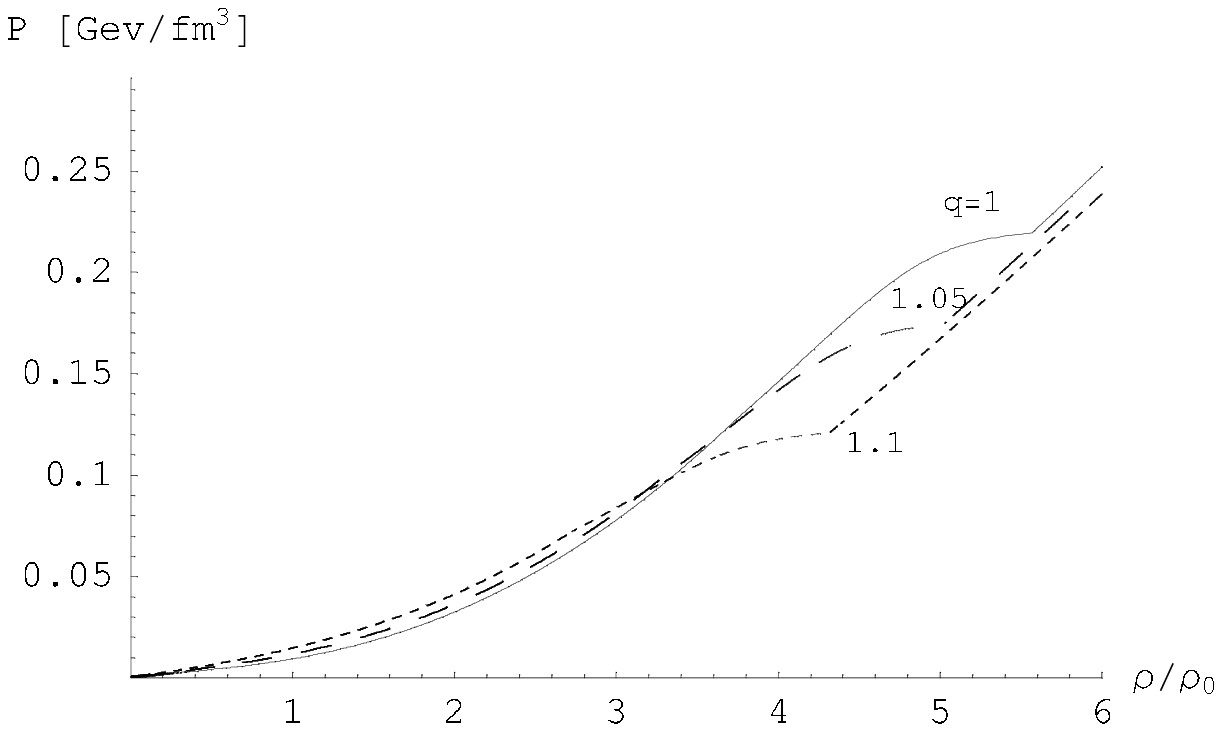}
} \caption{Pressure versus baryon density (in units of the nuclear
saturation density $\rho_0$) in the mixed hadron-quark phase for
different values of $q$. } \label{fig_MP3}
\end{figure}

In Fig. \ref{fig_MP1}, we report the pressure versus the baryon
chemical potential, in Fig. \ref{fig_MP2}, the pressure as a
function of the energy density and in Fig. \ref{fig_MP3} we report
the pressure versus baryon density (in units of the nuclear
saturation density $\rho_0$) in the mixed hadron-quark phase for
different values of $q$. For the hadronic phase we have used the
so-called GM2 set of parameters \cite{glen} and in the quark phase
the bag parameter is fixed to $B^{1/4}$=170 MeV. The temperature
is fixed at $T=60$ MeV and the proton fraction at $Z/A$=0.4,
physical values which are estimated to be realistic for high
energy heavy-ion collisions. The mixed hadron-quark phase starts
at $\rho=3.75\,\rho_0$ for $q=1$, at $\rho=3.31\,\rho_0$ for
$q=1.05$ and at $\rho=2.72\,\rho_0$ for $q=1.1$. It is important
to observe that for $q=1.1$ the second critical transition density
is also reached, separating the mixed phase from the pure
quark-gluon matter phase, at $\rho=4.29\,\rho_0$ while for
$q=1.05$ the second critical density is reached at
$\rho=5.0\,\rho_0$ and at $\rho=5.57\,\rho_0$ for $q=1$.

As a concluding remark we note that nonextensive statistical
effects become extremely relevant at large baryon density and
energy density, as the ones which can be reached in high energy
collisions experiments. This fact can be an important ingredient
in the realization of a hydrodynamic model as well as to obtain a
deeper microscopic connection with the experimental observable.

\end{document}